\documentclass[11pt,preprint]{aastex}

\newcommand{\etal}{{\it et~al.}}

\begin{document}

\title{Asteroid Diameters and Albedos from NEOWISE Reactivation Mission Years Six and Seven}

\author{Joseph R. Masiero\altaffilmark{1}, A.K. Mainzer\altaffilmark{2}, J.M. Bauer\altaffilmark{3}, R.M. Cutri\altaffilmark{1}, T. Grav\altaffilmark{2}, E. Kramer\altaffilmark{4}, J. Pittichov\'{a}\altaffilmark{4}, E.L. Wright\altaffilmark{5}}

\altaffiltext{1}{Caltech/IPAC, 1200 E California Blvd, MC 100-22, Pasadena, CA 91125, USA, {\it jmasiero@ipac.caltech.edu}}
\altaffiltext{2}{Lunar and Planetary Laboratory, University of Arizona, Tucson, AZ, 85721, USA}
\altaffiltext{3}{University of Maryland, College Park, MD, 20742, USA} 
\altaffiltext{4}{Jet Propulsion Laboratory/California Institute of Technology, Pasadena, CA 91109, USA}
\altaffiltext{5}{University of California, Los Angeles, CA, 90095, USA} 

\begin{abstract}

We present diameters and albedos computed for the near-Earth and Main
Belt asteroids observed by the Near-Earth Object Wide-field Infrared
Survey Explorer (NEOWISE) spacecraft during the sixth and seventh
years of its Reactivation mission.  These diameters and albedos are
calculated from fitting thermal models to NEOWISE observations of
$199$ NEOs and $5851$ MBAs detected during the sixth year of the
survey, and $175$ NEOs and $5861$ MBAs from the seventh year.
Comparisons of the near-Earth object diameters derived from
Reactivation data with those derived from the WISE cryogenic mission
data show a $\sim30\%$ relative uncertainty. This larger uncertainty
compared to data from the cryogenic mission is due to the need to
assume a beaming parameter for the fits to the shorter wavelength data
that the Reactivation mission is limited to.  We also present an
analysis of the orbital parameters of the Main Belt asteroids that
have been discovered by NEOWISE during Reactivation, finding that
these objects tend to be on orbits that result in their perihelia
being far from the ecliptic, and thus missed by other surveys.  To
date, the NEOWISE Reactivation survey has provided thermal fits of
$1415$ unique NEOs.  Including the mission phases before spacecraft
hibernation increases the count of unique NEOs characterized to $1845$
from WISE's launch to the present.

\end{abstract}

\section{Introduction}

Surveying in its Reactivated mission since 13 December 2013, the
Near-Earth Object Wide-field Infrared Survey Explorer (NEOWISE) has
obtained over 14 epochs of observations of the sky at thermal infrared
wavelengths.  Included in the archived, publicly-accessible data from
the survey are over one million photometric and astrometric
measurements of asteroids and comets.  These data provide a unique
resource for studying the small bodies of our Solar system, as well as
the time-variable infrared sky.

The NEOWISE mission \citep{mainzer14} makes use of the Wide-field
Infrared Survey Explorer (WISE) spacecraft \citep{wright10} which
follows a sun-synchronous polar orbit above the Earth's
  terminator, scanning rings around the sky over the course of each
orbit.  As the survey progresses, the same area of sky is repeatedly
imaged over the course of $\sim24~$hours, allowing moving objects to
be distinguished from background sources, and for the detection and
discovery of new near-Earth objects (NEOs).  Images and extracted
source catalogs from these data are archived and accessible at the
NASA/IPAC Infrared Science Archive
(IRSA\footnote{https://irsa.ipac.caltech.edu}).  The NEOWISE
Explanatory Supplement \citep{cutri15} provides a comprehensive
overview of the mission and data products.

The NEOWISE infrared observations provide measurements of the thermal
emission from a large number of small bodies in the inner Solar
system.  The data, when analyzed by means of thermal modeling
techniques, allow us to constrain the size of these bodies, as well as
their visible geometric albedo when ground-based visible light
measurements are available.  Previous works
\citep{nugent15,nugent16,masiero17,masiero20a} have presented the
results of thermal model fitting for NEOs and Main Belt asteroids
(MBAs) observed in the first 5 years of the NEOWISE reactivation.  In
this paper, we present results from Years 6 and 7 of the NEOWISE
Reactivation survey.  We also discuss some of the characteristics of
the population of MBAs discovered by NEOWISE during Reactivation, as
the extent of this population was somewhat unexpected.  Although
NEOWISE's sensitivity to objects in the Main Belt is greatly
diminished compared to the Cryogenic mission, it's unique field of
regard among asteroid surveys means it is more sensitive to different
subpopulations.

\section{Observations}

The NEOWISE survey pattern is fixed by the spacecraft's orbital
geometry.  Originally launched on an orbit perpendicular to the
Earth-Sun line, the slight atmospheric drag over the course of the
mission has reduced the spacecraft's altitude and caused the orbit to
precess, shifting NEOWISE's orbital plane by $\sim20^\circ$.  On the
side of increasing Solar elongation survey observations continue to be
made in a zenith-pointing direction (to reduce the thermal load on the
telescope), while on the side of decreasing elongation the survey
observations continue to point perpendicular to the Sun (due to the
constraints on pointing set by the sun shield) at the cost of an
increased heat load on the telescope from the Earth.

The detector performance has remained remarkably stable over the
course of the increased heating associated with orbital precession.
The telescope's cryostat acts as an insulator modulating the changes
in temperature over each orbit, though overall seasonal changes in
telescope temperature are observed following the precession of the
orbit.  These result in decreases to sensitivity at the $<0.02~$mag
level for W2 during the warmest phases of the orbit \citep[see Sec
  I.2.c.iii, Fig 11 and Sec IV.2.d.ii of][]{cutri15}.  Otherwise,
NEOWISE's recent performance has remained consistent with that of
previous survey phases.

NEOWISE scans an arc of sky each half-orbit. Every 11 seconds, the two
active $1024\times 1024$ pixel HgCdTe detectors capture an image of a
$47^\prime \times 47^\prime$ area of sky simultaneously at $3.4~\mu$m
and $4.6~\mu$m (hereafter, W1 and W2).  Each image has a $\sim10\%$
overlap with the previous image along the scan direction.  Source
detection and photometry are performed on each calibrated image,
providing a list of sources that are searched for candidate moving
objects.  The mission uses the WISE Moving Object Processing System
\citep[WMOPS,][]{mainzer11} to perform a blind search for Solar system
objects.  All candidate moving objects detected at least five times
and vetted to be real by automatic and human quality assurance are
reported the Minor Planet Center
(MPC)\footnote{https://www.minorplanetcenter.net} for publication and
archiving.  The MPC associates observations with known objects, or
assigns designations for newly discovered ones, as well as calculating
orbits for all minor planets.

The analysis presented here covers data obtained during the survey's
sixth and seventh years (13 Dec 2018 to 12 Dec 2019, and 13 Dec 2019
to 12 Dec 2020, respectively).  The NEOWISE Explanatory Supplement
\citep{cutri15}, which is updated for each data release, provides
further details about the mission parameters, data calibration, and
data access.  To extract data from the archive, we follow the
methodology described in previous NEOWISE-related publications
\citep[e.g.][]{masiero17}.  Physical properties from previous
publications are archived on the NASA Planetary Data System
\citep{mainzer19}, and the fits presented here will be added to that
archive at a future update.

\section{Thermal Modeling Technique}

Asteroids absorb incident sunlight, are warmed by it, and then re-emit
that energy as thermal emission.  When the orbit of an object is
sufficiently well-known the Solar flux it receives at any given time
is precisely constrained, and so a representational model can be used
to determine how the incident energy would be processed and emitted as
thermal radiation.  By comparing the modeled flux to the measured
infrared flux we can determine the size of the model that best
reproduces the measurements and therefore constrains the size of the
asteroid.  When rotation rates, rotation poles, and 3D asteroid shape
models are available along with multiple epochs of infrared data
detailed thermophysical models can be used to precisely constrain the
dimensions and thermal inertia of the body
\citep[e.g.][]{alilagoa14,koren15,yu17,hanus18}.  When resolved images
and {\it in situ} thermal measurements are available, maps of thermal
inertia across the surface can even be measured \citep{rozitis20}.

For the majority of objects observed by NEOWISE, little is known
beyond their orbit and a constraint on the absolute visible magnitude
$H_V$.  Therefore, we employ a simplified model to analyze their
thermal emission known as the Near-Earth Asteroid Thermal Model
\citep[NEATM,][]{harris98}.  NEATM assumes that the asteroid is a
sphere, with zero thermal emission contributed from the night side of
the body.  In this way, we can constrain the size of a sphere that
best fits the observed thermal flux and determine a
spherical-equivalent diameter for the body.  Real asteroids are not
spheres, though, so there is some inherent model uncertainty from the
shape of the body.  In addition, observations at larger Solar phase
angles will see larger fractions of the night side, increasing model
deviation from the actual size in a systematic way \citep[see][for a
  discussion on these uncertainties]{mommert18}.  In contrast, as
heliocentric distance and phase angle are coupled for NEOWISE due to
the survey pattern (with higher phase objects will be closer to the
Sun and warmer) the thermal emission for high phase objects will peak
closer to the W2 bandpass, which can result in smaller calculated
statistical uncertainties from the unknown beaming parameter.

NEATM employs a wavelength-independent variable $\eta$, known as the
beaming parameter, to allow the model to be adjusted to account for
the differences between the modeling assumptions and the actual
thermophysical properties of the surface \citep{harris98}.  This is a
simplification of actual asteroidal surfaces which can show emissivity
changes as a function of wavelength of order $10\%$
\citep{salisbury89}. The $\eta$ parameter is dependent on phase
\citep[e.g.][]{spencer89,harris07,delbo07,mainzer11neo}, though the
$\eta$-phase relation shows significant scatter due to a combination
of the specific observing conditions of the object as well as the
thermophysical properties of the surface
\citep{harris14,harris20}. Following previous work \citep{masiero20a}
the beaming parameter was held fixed for all fits as the majority of
cases have only a single thermally dominated band ($W2$).  That work
found that for the few cases that do have significant thermal emission
in $W1$, the reflected contribution to $W1$ is not well-enough known
to enable a strong constraint on the contribution of emission vs
reflection, leading to highly uncertain fits.  For the purposes of our
modeling, we assume $\eta=1.4\pm0.5$ for NEOs \citep{mainzer11neo},
while for MBAs we assume $\eta=0.95\pm0.2$ \citep{masiero11}.  The
uncertainty on $\eta$ is included in our Monte Carlo analysis to
propagate to the final uncertainty on the fitted parameters: diameter
and albedo. This uncertainty on assumed $\eta$ is significantly larger
than the change in expected $\eta$ across the phase angles our NEOs
are observed at, and thus accounts for any offset due to this
relationship.

Following the methods used for previous analyses of NEOWISE data
\citep[most recently][]{masiero20a}, we use the NEOWISE positions and
observations times reported to and archived by the Minor Planet Center
as the starting point for our data extraction.  This ensures that
observations have been vetted by both the NEOWISE mission during
processing and reporting as well as by the MPC during ingest.  We used
all observations in the years 6 and 7 time period with the C51
observatory code assigned to NEOWISE. These detections, along with the
associated spacecraft positions, were extracted from the MPC's
observation file and used as input for a query of the NEOWISE
Single-Exposure database hosted by IRSA to retrieve the photometry
associated with each detection.  The database was searched for
detections that matched the MPC-archived observations to within $5"$
in position and $5~$seconds in time.

The observations extracted from the NEOWISE database were filtered
based on the quality of the PSF-fit to the W2 band.  We focus on W2
for this quality cut because this band dominates the thermal fit for
NEOs and MBAs, and noise in this measurement can result in large
changes to the reported diameter.  We remove all detections from
consideration for fitting that have $w2rchi2>5$, which is the reduced
$\chi^2$ of the goodness of fit of the PSF to the source profile in
the W2 band.  This typically is a result of a cosmic ray strike
coincident with the detection, which would cause a dramatic increase
in the photometry and thus an over-estimate of diameter.  Additionally
we remove any objects from consideration for fitting that have orbital
arcs of less than $0.02~$years, indicating that these objects have not
received sufficient followup to have a well-constrained orbit.
Without a good orbit, the geocentric distance becomes
under-constrained, which has a negative impact on the quality of the
diameter fit.  

We also exclude detections that were coincident within the same search
radius with a stationary background object.  Background sources were
identified from the stacked Atlas catalog from the 2010 cryogenic WISE
survey data\footnote{ \it
https://irsa.ipac.caltech.edu/cgi-bin/Gator/nph-scan?mission=irsa\&submit=Select\&projshort=WISE
} \citep{cutri12}. Sources were only removed if $W2-W2_{atlas}>-1~$mag
to ensure that faint background objects would not remove valid
detections of bright asteroids.  This rejection of background
contaminated sources occurred for $2.9\%$ of sources in each survey
year.

Our implementation of NEATM uses a sphere made of 288 facets, with the
temperature on each facet and the flux observed from each facet
determined based on the observational geometry.  We color-correct the
flux from each facet based on its blackbody temperature following
\citet{wright10} with corrections ranging from 1.005 for reflected
light to 1.1-1.5 for emission from subsolar facets (with cooler facets
having much larger corrections but significantly smaller contributions
to the total flux).  The total modeled flux is then compared to the
measured flux, and the variable parameters diameter ($D$) and visible
geometric albedo ($p_V$) are found through the least-squares
optimization routine in the {\it scipy} package \citep{scipy}.  To
reduce diameter bias due to the unknown shape and rotation phase at
the time of observation, we require a minimum of three detections with
W2 photometric uncertainties $w2sigmpro<0.25~$mag, which are fit
simultaneously with a single spherical model.  It should be noted that
to be identified as a candidate by the WMOPS system an object must be
detected at least five times, so the three detection minimum would
only reject objects for cases where a cosmic ray or nearby star
spoiled multiple detections.  This only occurred for 8 NEOs and 75
MBAs across both years of data.

To constrain the albedo, we require an external measurement of the
visible brightness of the body.  For this work, we use the published
absolute visible magnitude ($H_V$) included in the MPC orbital
catalog, which is calculated by the MPC (along with the $G_V$ slope
parameter) based on all submitted photometry after conversion to $V$
band.  The uncertainty on $H_V$ is a primary component of the
uncertainty on the derived albedo for NEOs \citep[see][for a
  discussion of this uncertainty]{masiero21}, along with the
uncertainty on diameter from the unknown beaming parameter value which
is often the dominant component in these cases.  However, uncertainty
measurements are not included with the MPC's $H_V$ values, and so must
be assumed for the vast majority of objects we fit here.  Previous
research \citep[e.g.][]{pravec12,veres15} has highlighted systematic
biases in the MPC $H_V$ magnitude values for subsets of the data.
Following previous work \citep[e.g.][]{masiero20a}, we update our
$H_V$ magnitudes to the improved values from \citet{veres15} and
assume an uncertainty of $0.2~$mag on the $H_V$ for near-Earth
objects, and $0.05~$mag for Main Belt asteroids as these tend to have
a significant number of archived visible light observations.  To model
the reflected light component we assume that the $G_V$ slope parameter
has an uncertainty of $0.1$, and that the ratio of albedos in the
infrared to the visible is $p_{IR}/p_V=1.6\pm1.0$ for NEOs and
$p_{IR}/p_V=1.5\pm0.5$ for MBAs following previous analyses
\citep{mainzer11neo,masiero11}. Once formal uncertainties are
available for absolute magnitude measurements for the majority of
asteroids, characterization of the uncertainty on albedo will improve
somewhat.

The diameter constraints determined by fitting NEATM models to the
NEOWISE data are only valid for cases where at least one band is
dominated by thermal emission.  As noted in previous work
\citep[e.g.][]{masiero20a}, due to the uncertainty on the ratio of
IR-to-visible albedos NEATM fits with reflected light contributions
$>10\%$ enter a regime where the model becomes under-constrained and
highly uncertain.  We check the reflected light contribution in each
band for every fit, and remove any with $>10\%$ reflected light
contribution to W2. This resulted in the removal of $29\%$ of MBAs in
both years, $8\%$ of NEOs observed in Year 6, and $13\%$ of NEOs
observed in Year 7.  A detailed description of the reflected light
model can be found in \citet{mainzer11cal}.

To assess the statistical uncertainty on the resulting diameter and
albedo fits, we perform Monte Carlo trials using the uncertainties on
the measured NEOWISE and $H_V$ magnitudes, as well as the defined
variation ranges on the assumed parameters ($\eta$, $p_{IR}/p_V$).  We
perform 25 trials, with each trial drawing a different measured or
assumed parameter using a normal distribution with the initial value
as the mean and the uncertainty as the width.  The standard deviation
of the output fitted values ($D$ and $p_V$) is taken as the
statistical uncertainty for the fit of those values.  There are
additional systematic uncertainties inherent to our fitting method
that will also affect our results, which we discuss below.

\section{Results}

\begin{figure}[ht]
\begin{center}
\includegraphics[scale=0.6]{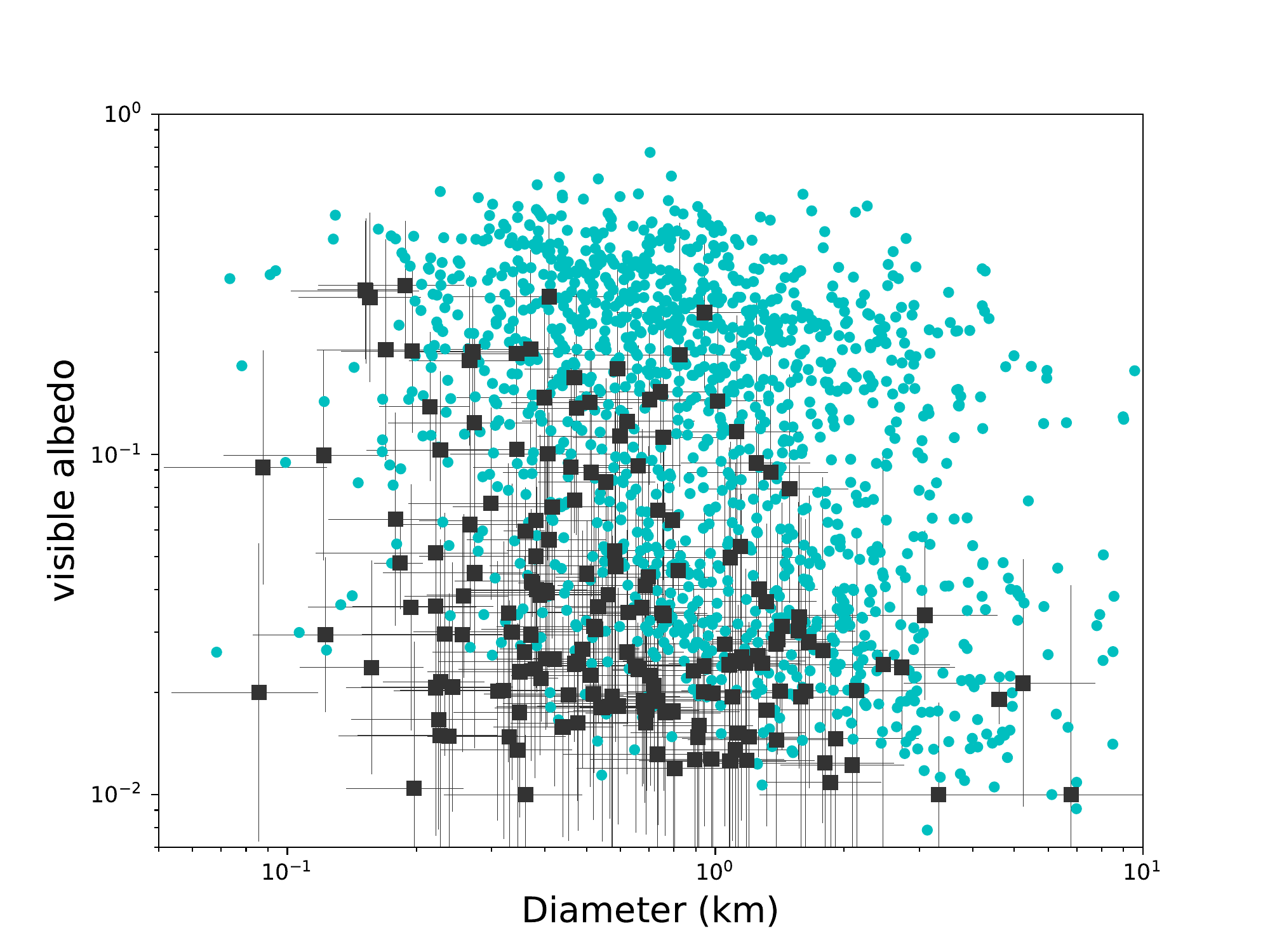}
\protect\caption{Diameters derived from thermal modeling compared to
  geometric visible albedos for all objects observed by NEOWISE (cyan
  circles) and those discovered by NEOWISE (black squares) during the
  first seven years of the Reactivation mission.  Objects discovered
  by NEOWISE tend to be lower albedo, and larger than a few hundred
  meters, occupying a niche that other surveys are less successful at
  detecting.  Uncertainties are only shown for discovered objects for
  clarity, but are comparable in size for the previously known
  asteroids.}
\label{fig.diamalb}
\end{center}
\end{figure}

We present the results of our diameter and albedo fitting in tables
below.  The NEOs and MBAs observed in Year 6 are given in
Table~\ref{tab.neo6} and Table~\ref{tab.mba6}, respectively, while the
NEO and MBA fits from Year 7 are given in Table~\ref{tab.neo7} and
Table~\ref{tab.mba7}, respectively.  The Year 6 table contains
physical property measurements for 199 NEOs and 5851 MBAs, while the
Year 7 table contains 175 NEOs and 5861 MBAs.  Each table lists the
MPC-packed designation for the object; the $H_V$ and $G_V$ absolute
magnitude and slope parameters that were used as input measurements
for the fits; the diameter and albedo with associated statistical
errors; the number of detections in the $W1$ and $W2$ bands used for
the fit; and the phase angle at the midpoint of observation. Fitted
albedos are presented with asymmetric error bars as the uncertainties
are derived from magnitude uncertainties, which are symmetric in
log-space, and to ensure the low albedo objects with
poorly-constrained $H_V$ values won't have albedo measurements smaller
than the $1-\sigma$ uncertainty. Figure~\ref{fig.diamalb} shows the
diameter and albedo fits for all NEOs observed during the first seven
years of the Reactivation survey, combining these results with those
from previous publications \citep{mainzer19,masiero20a}.

Combining the model output diameter and albedo values will allow for
the determination of the 'best-fit' $H_V$ that resulted from the joint
emitted light/reflected light model.  In cases where $W1$ is dominated
by reflection, the reflected light component is fit simultaneously to
both the measured $W1$ and predicted $V$ fluxes based on the input
$H_V$ and $G_V$. If the uncertainty on the $W1$ measurement is smaller
than the uncertainty on $H_V$ (which is often the case) this will
result in the model fit favoring $W1$ and the output $H_V$ deviating
from the input $H_V$ within the uncertainty.  These differences show
systematic offsets at the $\sim0.2~$mag level because: 1) the
$p_{IR}/p_V$ ratio depends on the composition of asteroid
\citep{mainzer11tax} while we assume a single median value for the
whole population; 2) our removal of objects with significant reflected
light in $W2$ preferentially removes high-albedo objects like S-types
\citep{masiero17} from the Main Belt dataset skewing the expected
albedo ratio; and 3) the MPC's $H_V$ determinations are known to have
systematic offsets \citep{pravec12,veres15}.

Objects will appear multiple times within a single results table when
they have been seen at multiple different epochs within that year.
When the orbital and viewing geometries are sufficiently different
between observations a simple NEATM fit cannot be used across epochs
due to changes in subsolar temperature and phase angle, so these
objects are presented as separate entries for each epoch.  Differences
in fitted diameter between epochs are a combination of the changing
viewing aspect of the non-spherical body (and thus change to the
effective spherical diameter fit to the projected shape) along with
statistical and model uncertainties.  When multiple epochs are
available, along with ancillary data such as rotation period and pole,
more advanced thermophysical modeling can provide insights into the 3D
shape and thermophysical properties of the surface \citep[for example,
  see][]{koren15,hanus16,hanus18,masiero19}.  Large differences
between the fits from different epochs presented here indicate objects
that would likely benefit from more advanced modeling.

One object in the Year 6 NEO list has an anomalously low albedo
($p_V=0.005$) and a very large diameter (D$=49\pm32~$km), even after
filtering of the data and verification of the detections.  This
object, (3552) Don Quixote, has a $6.665~$ hr period and a light curve
amplitude that can be as large as $1.24~$mag \citep{warner19},
indicating it is highly elongated.  The half-period of this object's
rotation is nearly the same as the $\sim188~$min period between
NEOWISE observations, as determined by the orbit and survey plan,
meaning that we expect significant aliasing in the observations.  If
NEOWISE detected this object near a light curve maximum, we would
expect all observations from that epoch to be near the maximum,
resulting in a size constraint that reflects only the longest axis.
Using an $H_V$ value derived from all rotation phases would then
result in an anomalously low albedo.

Alternately, Don Quixote is known to be an active, cometary object
\citep{mommert14}, and emission of gas and dust will cause the
measured fluxed in W1 and W2 to be increased above what would be
expected for the bare nucleus.  Emission from CO and CO$_2$ in
particular can cause significant flux excesses in the W2 band
\citep[e.g.][]{bauer15}.  While our images show no discernible coma or
tail, recent observations of Don Quixote have shown that it
experienced cometary outgassing approximately 6 months before the
NEOWISE observations \citep{mommert20}, and any coma contribution to
the observed flux from dust or gas would also result in an increase in
the derived size and thus the very low albedo constraint.  Our size
and albedo constraints derived here for (3552) Don Quixote would be
expected to be less reliable than for a typical object, which is
reflected in the very large uncertainty on the diameter. (3552) Don
Quixote was also observed by NEOWISE in Year 5 of its survey before
the onset of the most recent activity providing a diameter constraint
of $27\pm13~$km and an albedo of $0.03^{+0.03}_{-0.02}$
\citep{masiero20a}, more consistent with other literature values of
$D=19~$km \citep{chapman94} and $18.4~$km \citep{mommert14}.

Two objects in the Year 7 NEO list also yield anomalously low albedos
in our fits: (248590) and 2008 CG119.  NEO (248590) has three other
epochs of fitted diameters in the NEOWISE data \citep{mainzer19}, with
reported albedos of $p_V=3.7\%, 1.8\%,$ and $1.1\%$.  This, along with
the albedo here of $p_V=0.9\%$, points to a likely incorrect $H_V$
magnitude, which would bias all of the albedos to very low values.
Fitting $H_V$ magnitudes suffers from numerous potential sources of
error, and so large offsets can occur, especially for NEOs
\citep{masiero21}.  In this case, our assumed uncertainty on the $H_V$
magnitude may also be underestimated.  In contrast, 2008 CG119 only
had three detections that survived our quality cuts, increasing the
chances that rotational light curve variations would bias the diameter
fit upward, and thus the albedo downward.  In addition, the
uncertainty on the diameter fit is large, further complicating
interpretation of the albedo.

\begin{table}[ht]
\begin{center}
\caption{Thermal model fits for NEOs detected in the sixth year of the
  NEOWISE survey.  Table 1 is published in its entirety
  in the electronic edition; a portion is shown here for guidance
  regarding its form and content.}
\vspace{1ex}
\noindent
\begin{tabular}{ccccccccc}
  \tableline
  Name  &  H$^\dagger$  &   G   &   Diameter  &  $p_V~^{\dagger\dagger}$  &  beaming & n$_{W1}$  &n$_{W2}$ & phase  \\
  & (mag) & & (km) & & & & & (deg)  \\
  \tableline
  02059 & 15.90 &  0.15 &    1.84 $\pm$   0.79 & 0.179 (+0.188/-0.092) & 1.40 $\pm$ 0.50 &  20 &  20 & 46.50 \\
  02100 & 16.30 &  0.15 &    2.05 $\pm$   0.83 & 0.126 (+0.123/-0.062) & 1.40 $\pm$ 0.50 &   0 &   5 & 58.48 \\
  02100 & 16.30 &  0.15 &    1.46 $\pm$   0.52 & 0.251 (+0.263/-0.128) & 1.40 $\pm$ 0.50 &   0 &  13 & 66.45 \\
  03102 & 16.20 &  0.15 &    1.80 $\pm$   0.64 & 0.241 (+0.213/-0.113) & 1.40 $\pm$ 0.50 &  23 &  25 & 53.37 \\
  03200 & 14.30 &  0.15 &    2.95 $\pm$   1.12 & 0.194 (+0.182/-0.094) & 1.40 $\pm$ 0.50 &   7 &   7 & 46.02 \\
  03552 & 12.90 &  0.15 &   49.02 $\pm$  31.66 & 0.005 (+0.009/-0.003) & 1.40 $\pm$ 0.50 &   0 &  12 & 18.79 \\
  04487 & 17.30 &  0.15 &    1.24 $\pm$   0.41 & 0.138 (+0.181/-0.078) & 1.40 $\pm$ 0.50 &   0 &  12 & 51.96 \\
  05189 & 17.80 &  0.15 &    0.55 $\pm$   0.17 & 0.448 (+0.323/-0.188) & 1.40 $\pm$ 0.50 &   0 &   4 & 69.10 \\
  05626 & 14.20 &  0.15 &    4.78 $\pm$   1.99 & 0.181 (+0.182/-0.091) & 1.40 $\pm$ 0.50 &  21 &  23 & 52.08 \\
  06178 & 15.80 &  0.15 &    4.22 $\pm$   1.77 & 0.047 (+0.071/-0.028) & 1.40 $\pm$ 0.50 &   0 &  13 & 41.18 \\

\hline
\end{tabular}
\label{tab.neo6}
$^\dagger$Measured H used as input for the modeling; the model-output H value can be found using the output diameter, albedo, and the equation $D = 1329\times10^{H/-5}/\sqrt{p_V}$\\
$^{\dagger\dagger}$Albedo uncertainties are symmetric in log-space due to the uncertainty on $H$; the asymmetric linear equivalents of the $1 \sigma$ log-space uncertainties are presented here.
\end{center}
\end{table}

\begin{table}[ht]
\begin{center}
\caption{Thermal model fits for MBAs detected in the sixth year of the
  NEOWISE survey.  Table 2 is published in its entirety
  in the electronic edition; a portion is shown here for guidance
  regarding its form and content.}
\vspace{1ex}
\noindent
\begin{tabular}{ccccccccc}
  \tableline
  Name  &  H$^\dagger$  &   G   &   Diameter  &  $p_V~^{\dagger\dagger}$  &  beaming & n$_{W1}$  &n$_{W2}$ & phase  \\
  & (mag) & & (km) & & & & & (deg)  \\
  \tableline
  00021 &  7.50 &  0.15 &  108.04 $\pm$  28.59 & 0.131 (+0.079/-0.049) & 0.95 $\pm$ 0.20 &   6 &   6 & 24.75 \\
  00034 &  8.60 &  0.15 &  116.91 $\pm$  35.99 & 0.036 (+0.054/-0.021) & 0.95 $\pm$ 0.20 &  12 &   8 & 20.48 \\
  00036 &  8.60 &  0.15 &  106.25 $\pm$  25.32 & 0.047 (+0.029/-0.018) & 0.95 $\pm$ 0.20 &  13 &  13 & 24.83 \\
  00038 &  8.50 &  0.15 &   86.66 $\pm$  23.47 & 0.094 (+0.058/-0.036) & 0.95 $\pm$ 0.20 &   0 &  10 & 23.05 \\
  00045 &  7.50 &  0.15 &  230.72 $\pm$  89.48 & 0.033 (+0.051/-0.020) & 0.95 $\pm$ 0.20 &   0 &   7 & 20.30 \\
  00045 &  7.50 &  0.15 &  220.00 $\pm$  63.65 & 0.041 (+0.027/-0.016) & 0.95 $\pm$ 0.20 &  12 &  10 & 23.24 \\
  00046 &  8.50 &  0.15 &  132.76 $\pm$  60.42 & 0.042 (+0.047/-0.022) & 0.95 $\pm$ 0.20 &  10 &  10 & 20.55 \\
  00046 &  8.50 &  0.15 &  116.80 $\pm$  39.19 & 0.047 (+0.037/-0.021) & 0.95 $\pm$ 0.20 &  12 &  10 & 24.06 \\
  00051 &  7.60 &  0.15 &  138.85 $\pm$  36.79 & 0.075 (+0.045/-0.028) & 0.95 $\pm$ 0.20 &  10 &   9 & 21.78 \\
  00051 &  7.60 &  0.15 &  126.55 $\pm$  34.76 & 0.088 (+0.057/-0.035) & 0.95 $\pm$ 0.20 &  14 &  15 & 24.85 \\
\hline
\end{tabular}
\label{tab.mba6}

$^\dagger$Measured H used as input for the modeling; the model-output H value can be found using the output diameter, albedo, and the equation $D = 1329\times10^{H/-5}/\sqrt{p_V}$\\
$^{\dagger\dagger}$Albedo uncertainties are symmetric in log-space due to the uncertainty on $H$; the asymmetric linear equivalents of the $1 \sigma$ log-space uncertainties are presented here.

\end{center}
\end{table}

\begin{table}[ht]
\begin{center}
\caption{Thermal model fits for NEOs detected in the seventh year of the
  NEOWISE survey.  Table 3 is published in its entirety
  in the electronic edition; a portion is shown here for guidance
  regarding its form and content.}
\vspace{1ex}
\noindent
\begin{tabular}{ccccccccc}
  \tableline
  Name  &  H$^\dagger$  &   G   &   Diameter  &  $p_V~^{\dagger\dagger}$  &  beaming & n$_{W1}$  &n$_{W2}$ & phase  \\
  & (mag) & & (km) & & & & & (deg)  \\
  \tableline
  00887 & 13.78 &  0.15 &    6.62 $\pm$   2.04 & 0.124 (+0.151/-0.068) & 1.40 $\pm$ 0.50 &  10 &   9 & 37.41 \\
  01685 & 14.31 &  0.15 &    2.61 $\pm$   0.72 & 0.395 (+0.247/-0.152) & 1.40 $\pm$ 0.50 &  12 &  13 & 70.62 \\
  02102 & 16.01 &  0.15 &    1.43 $\pm$   0.52 & 0.308 (+0.267/-0.143) & 1.40 $\pm$ 0.50 &   9 &   9 & 52.93 \\
  03360 & 15.91 &  0.15 &    2.70 $\pm$   1.01 & 0.138 (+0.123/-0.065) & 1.40 $\pm$ 0.50 &   5 &   5 & 56.89 \\
  03360 & 15.91 &  0.15 &    3.17 $\pm$   1.76 & 0.076 (+0.155/-0.051) & 1.40 $\pm$ 0.50 &   0 &  10 & 35.06 \\
  04055 & 14.71 &  0.15 &    2.61 $\pm$   1.03 & 0.335 (+0.317/-0.163) & 1.40 $\pm$ 0.50 &  37 &  38 & 55.59 \\
  04953 & 14.99 &  0.15 &    2.73 $\pm$   1.00 & 0.228 (+0.197/-0.106) & 1.40 $\pm$ 0.50 &   5 &   6 & 50.56 \\
  07753 & 17.96 &  0.15 &    1.64 $\pm$   0.71 & 0.043 (+0.075/-0.027) & 1.40 $\pm$ 0.50 &   0 &  10 & 39.21 \\
  08014 & 18.90 &  0.15 &    0.34 $\pm$   0.11 & 0.433 (+0.339/-0.190) & 1.40 $\pm$ 0.50 &   0 &   6 & 75.94 \\
  09172 & 16.32 &  0.15 &    3.88 $\pm$   1.87 & 0.035 (+0.088/-0.025) & 1.40 $\pm$ 0.50 &   0 &   6 & 38.96 \\
\hline
\end{tabular}
\label{tab.neo7}

$^\dagger$Measured H used as input for the modeling; the model-output H value can be found using the output diameter, albedo, and the equation $D = 1329\times10^{H/-5}/\sqrt{p_V}$\\
$^{\dagger\dagger}$Albedo uncertainties are symmetric in log-space due to the uncertainty on $H$; the asymmetric linear equivalents of the $1 \sigma$ log-space uncertainties are presented here.

\end{center}
\end{table}

\begin{table}[ht]
\begin{center}
\caption{Thermal model fits for MBAs detected in the seventh year of the
  NEOWISE survey.  Table 4 is published in its entirety
  in the electronic edition; a portion is shown here for guidance
  regarding its form and content.}
\vspace{1ex}
\noindent
\begin{tabular}{ccccccccc}
  \tableline
  Name  &  H$^\dagger$  &   G   &   Diameter  &  $p_V~^{\dagger\dagger}$  &  beaming & n$_{W1}$  &n$_{W2}$ & phase  \\
  & (mag) & & (km) & & & & & (deg)  \\
  \tableline
  00013 &  6.79 &  0.15 &  228.76 $\pm$  71.61 & 0.043 (+0.043/-0.021) & 0.95 $\pm$ 0.20 &   7 &   8 & 24.22 \\
  00034 &  8.63 &  0.15 &  100.60 $\pm$  29.01 & 0.042 (+0.045/-0.022) & 0.95 $\pm$ 0.20 &   8 &   7 & 20.59 \\
  00038 &  8.54 &  0.15 &  113.40 $\pm$  37.72 & 0.044 (+0.036/-0.020) & 0.95 $\pm$ 0.20 &  15 &  14 & 22.20 \\
  00049 &  7.84 &  0.15 &  151.38 $\pm$  42.01 & 0.041 (+0.040/-0.020) & 0.95 $\pm$ 0.20 &   9 &   7 & 21.73 \\
  00049 &  7.84 &  0.15 &  138.89 $\pm$  46.07 & 0.049 (+0.038/-0.021) & 0.95 $\pm$ 0.20 &  26 &  24 & 21.96 \\
  00056 &  8.52 &  0.15 &  130.75 $\pm$  28.65 & 0.040 (+0.020/-0.013) & 0.95 $\pm$ 0.20 &  11 &  15 & 27.27 \\
  00058 &  9.04 &  0.15 &   87.34 $\pm$  27.34 & 0.045 (+0.037/-0.020) & 0.95 $\pm$ 0.20 &  11 &  10 & 21.09 \\
  00058 &  9.04 &  0.15 &   93.52 $\pm$  30.99 & 0.039 (+0.030/-0.017) & 0.95 $\pm$ 0.20 &   5 &   5 & 22.38 \\
  00066 &  9.54 &  0.15 &   67.89 $\pm$  27.57 & 0.042 (+0.056/-0.024) & 0.95 $\pm$ 0.20 &  19 &  20 & 19.30 \\
  00072 &  9.10 &  0.15 &   68.27 $\pm$  18.68 & 0.091 (+0.069/-0.039) & 0.95 $\pm$ 0.20 &  12 &  11 & 24.37 \\
\hline
\end{tabular}
\label{tab.mba7}

$^\dagger$Measured H used as input for the modeling; the model-output H value can be found using the output diameter, albedo, and the equation $D = 1329\times10^{H/-5}/\sqrt{p_V}$\\
$^{\dagger\dagger}$Albedo uncertainties are symmetric in log-space due to the uncertainty on $H$; the asymmetric linear equivalents of the $1 \sigma$ log-space uncertainties are presented here.

\end{center}
\end{table}

\begin{figure}[ht]
\begin{center}
\includegraphics[scale=0.6]{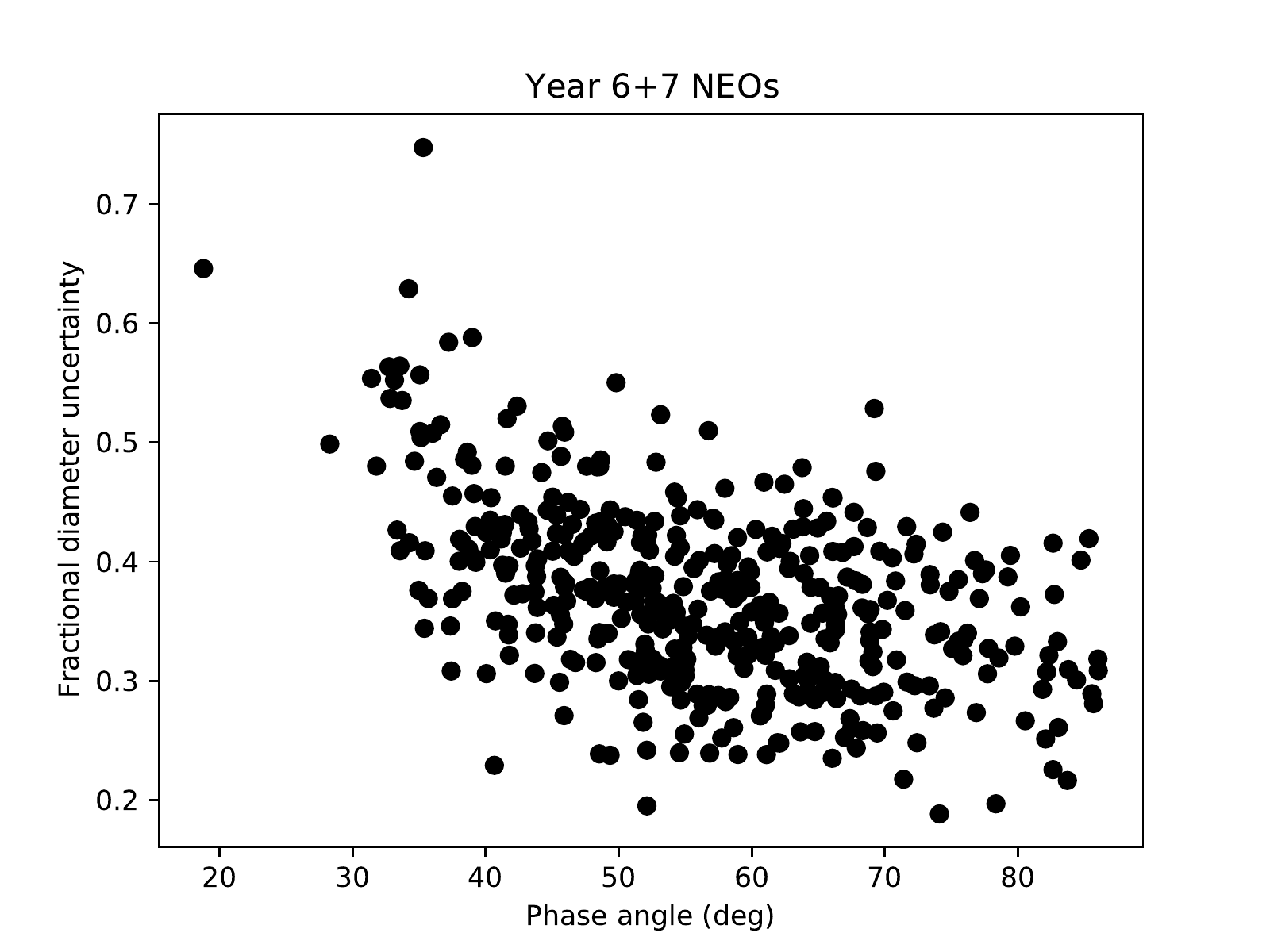}
\protect\caption{Fractional diameter uncertainty ($\sigma_D/D$) for
  all NEO measurements in NEOWISE Years 6 and 7 vs phase angle.  The
  uncertainty shown here only includes the statistical component
  determined through our Monte Carlo analysis.  Objects at smaller
  phase angles show larger fractional uncertainties because the peak
  of thermal emission is further from the W2 bandpass, providing a
  weaker constraint on the size given the large range of possible
  beaming values.}
\label{fig.uncphase}
\end{center}
\end{figure}

As discussed above, phase angle is linked to heliocentric distance and
thus temperature for NEOWISE, with higher phase objects being warmer.
Given the large range of possible beaming values for NEOs, cooler
objects would be expected to show larger statistical diameter
uncertainties as the peak flux will be further from the bandpass. This
is shown in Figure \ref{fig.uncphase} where the NEOs at phases below
$\alpha=40^\circ$ show increased statistical diameter
uncertainty. This is the opposite behavior as would be expected for
the systematic uncertainty, as described by
\citet{mommert18}. Systematic uncertainty components must be derived
on a population level by comparison with other diameter datasets (as
described in the next section).

\clearpage

\section{Discussion}

\subsection{Diameter Accuracy}

The accuracy of the diameter fits from the NEOWISE Reactivation survey
has been highly stable over the lifetime of the mission
\citep[cf.][]{masiero20a}.  This is not surprising given that the
photometric performance of the system has been nearly unchanged since
launch \citep{cutri15} and the thermal modeling methodology has also
remained unchanged.  Previous works have demonstrated this primarily
through comparison of diameter fits for Main Belt asteroids due to the
much larger number of MBAs observed by the mission.  This is because
MBAs are more numerous, and because the synodic periods of NEOs are
longer, and so opportunities to re-observe them are less frequent.

However, now that NEOWISE has been operating in its Reactivation
mission for over 7 years, we have a sufficiently large sample of
observed NEOs that also have literature diameter measurements from the
cryogenic NEOWISE mission, as well as a smaller set with diameter
constraints independent of thermal infrared models such as asteroid
radar echos or occultations.  Comparisons between the Reactivation
data and the cryogenic NEOWISE diameters allow us to determine the
effect on the fits of only fitting the Wien's side of the blackbody
emission spectrum (compared to fitting near the peak of the spectrum
with the cryogenic data), while comparison with non-IR diameters allow
us to assess the accuracy of the overall NEATM method.

\begin{figure}[ht]
\begin{center}
\includegraphics[scale=0.5]{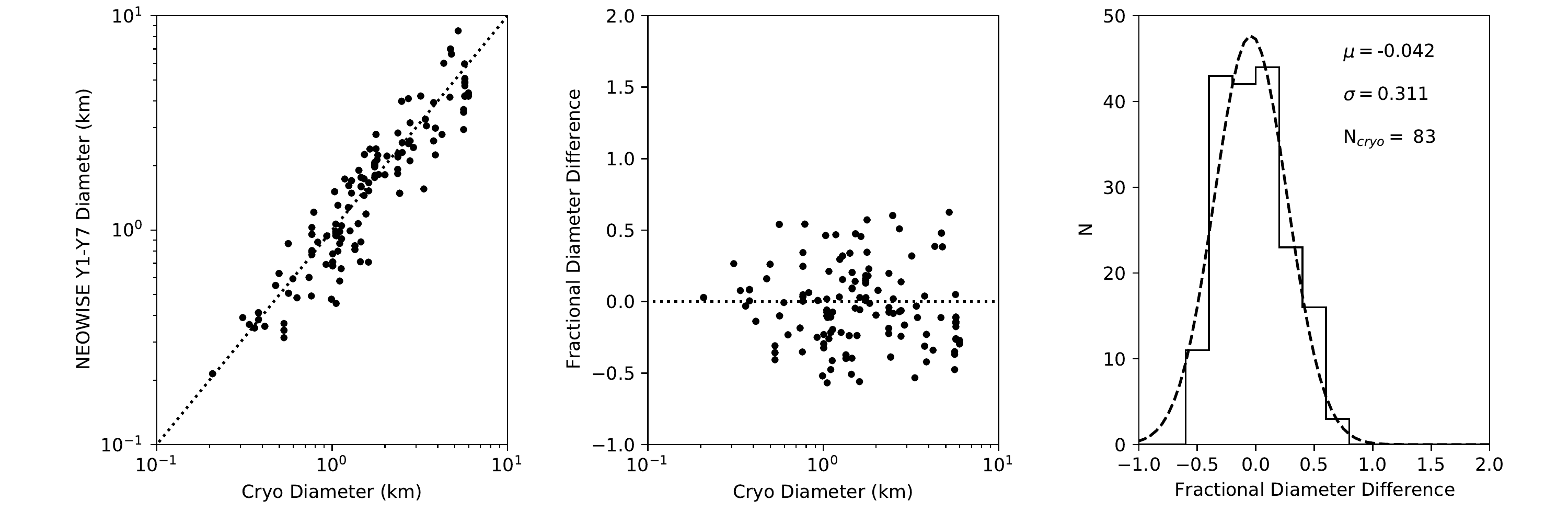}
\protect\caption{Diameters of NEOs observed during the first seven
  years of Reactivation compared to the diameter fits based on
  cryogenic NEOWISE data (left panel, with the dotted line indicating
  a one-to-one relationship).  The center panel shows the fractional
  diameter difference of the fits compared to the cryogenic diameters,
  while the right panel shows a histogram of the fractional diameter
  differences as well as a best-fit Gaussian distribution to the
  histogram (dashed line, with the Gaussian mean ($\mu$) and standard
  deviation ($\sigma$) given).}
\label{fig.cryocomp}
\end{center}
\end{figure}

NEOWISE has observed 83 unique NEOs at 182 epochs during the first
seven years of Reactivation that also have diameter fits from the
cryogenic phase of the NEOWISE mission \citep{mainzer11neo}.  The
comparison between the cryo fits and the Reactivation fits is shown in
Figure~\ref{fig.cryocomp}.  The best fit Gaussian to the fractional
diameter differences (Reactivation diameter minus the cryo diameter
divided by the cryo diameter) indicates that there is a slight
systematic offset between the datasets of $4\%$, with the cryo
diameters being larger.  The $1-\sigma$ scatter of $31\%$ is
attributed to a combination of the different viewing geometries
presenting different projections of the profile, as well as the need
to assume a beaming parameter for all of the Reactivation fits.

\begin{figure}[ht]
\begin{center}
\includegraphics[scale=0.5]{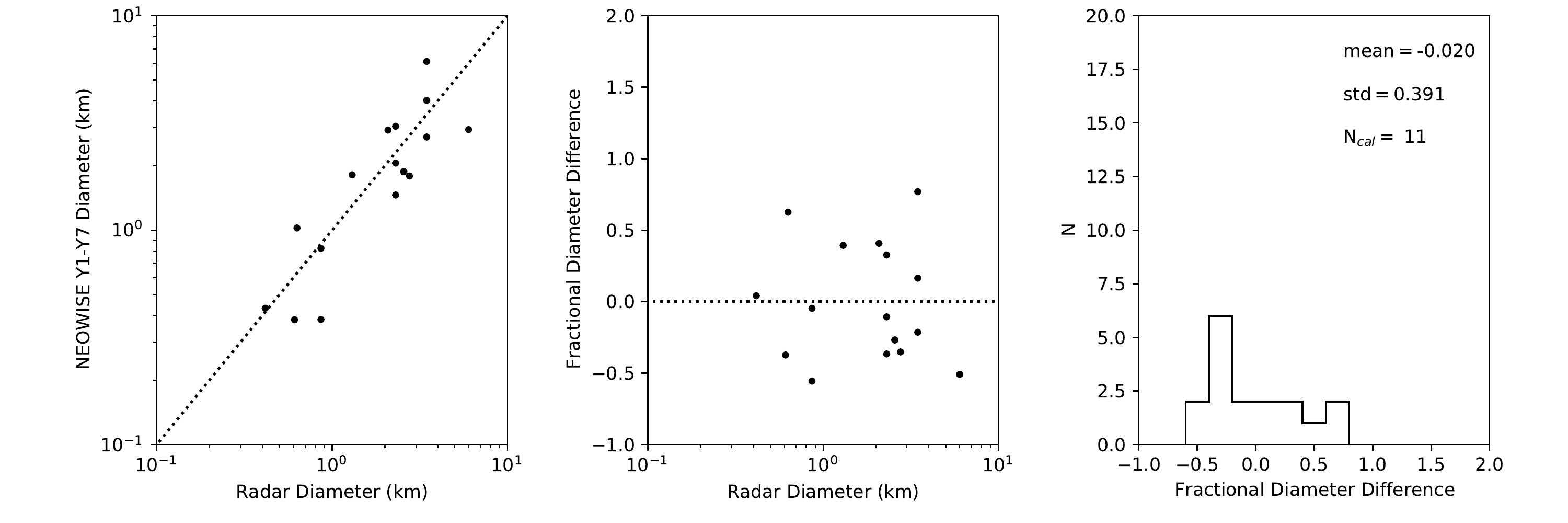}
\protect\caption{The same as Fig~\ref{fig.cryocomp} but showing the
  comparison between diameters from NEOWISE Reactivation and sizes
  from 3D models of radar observations. Vertical alignments of points
  in the first two panels are due to individual NEOs observed at
  multiple NEOWISE epochs.  As there is insufficient data for a
  reliable Gaussian fit, we instead present the mean and standard
  deviation of the fractional diameter difference distribution.}
\label{fig.radcomp}
\end{center}
\end{figure}

We also compare our Reactivation NEO fits to the spherical equivalent
diameters derived from 3D modeling of radar echoes of NEOs obtained at
the Arecibo and Goldstone facilities \citep{benner15}, as shown in
Figure~\ref{fig.radcomp}.  Radar diameters are obtained from a set of
publications presenting individual or collections of fits
\citep{hudson95, hudson99, benner99, shepard06, busch07, busch08,
  jimenez10, magri11, naidu15, taylor19, lawrence20}. Because of the
difficulty in obtaining sufficiently high signal-to-noise radar
observations to allow for 3D modeling, there are only 11 NEOs in the
overlap between the two datasets [(1620) Geographos, (2063) Bacchus,
  (2100) Ra-Shalom, (3200) Phaethon, (4179) Toutatis, (8567) 1996 HW1,
  (29075) 1950 DA, (33342) 1998 WT24 (185851) 2000 DP107, (276049)
  2002 CE26, (363067) 2000 CO101], representing 16 epochs of NEOWISE
observation.  Having a 3D radar model is important, however, as this
typically results in a better measurement of the asteroid's shape and
spherical equivalent size.  Due to the limited sample size, a Gaussian
fit would not be sufficiently constrained, so instead we calculate the
mean and standard deviation of the fractional diameter differences.
From the limited data we can see that the systematic offset in the
sizes is comparable to that seen for the comparison with the cryogenic
NEOWISE data, while the scatter is somewhat larger ($\sim39\%$).  This
may be a result of the combination of the effects described above as
well as the differences between NEATM model and the radar modeling
techniques which are $10-15\%$ when the beaming parameter can be
fitted to multiple infrared wavelengths \citep{mainzer11cal}, or may
be simply due to small number statistics.  As NEOWISE continues to
collect data, and more radar shape models are published, the sample
size should improve and allow for better determination of the
differences between model results.

\subsection{Objects in the Main Belt Discovered by NEOWISE During Reactivation}

As an infrared survey, the detection efficiency of NEOWISE is a
function of the temperature of the object.  Objects closer to the Sun,
and thus warmer, will be significantly brighter than an object of
similar size further from the Sun even after correcting for
$\frac{1}{r^2}$ distance effects.  Thus the $40~$cm diameter primary
mirror used by NEOWISE has comparable sensitivity to much larger
visible-light telescopes on the ground when it comes to detecting
asteroids and comets near the Earth.

For more distant objects like Main Belt asteroids the peak of the
thermal emission moves to longer wavelengths, decreasing the W2
signal-to-noise, while the W1 signal is almost entirely due to
reflected light.  For these objects NEOWISE's sensitivity is
comparable to a ground-based visible light telescope of similar size.
While NEOWISE discovered tens of thousands of Main Belt asteroids
during its cryogenic mission \citep{masiero11}, after the loss of
cryogen we would expect the number of Main Belt discoveries to be
minimal.

Instead, we find that a small number of objects in the Main Belt are
still being discovered by NEOWISE.  To investigate these objects, and
the reason that they were not previously found by other ground-based
surveys, we searched the MPC's Observation Archive\footnote{\it
https://minorplanetcenter.net/iau/ECS/MPCAT-OBS/MPCAT-OBS.html} for
all NEOWISE observations of Main Belt asteroids from 2013 onward that
received a `*' character designating a discovery observation for an
object.  For some objects multiple observations will receive a
discovery designation by the MPC, in cases where tracklets from
different years were given provisional designations and then later
linked.  We take an inclusive view of these when a NEOWISE discovery
observation was not the first discovery detection, as these indicate
cases where the previous observations resulted in an orbit
sufficiently uncertain as to essentially be lost.

\begin{figure}[ht]
\begin{center}
\includegraphics[scale=0.6]{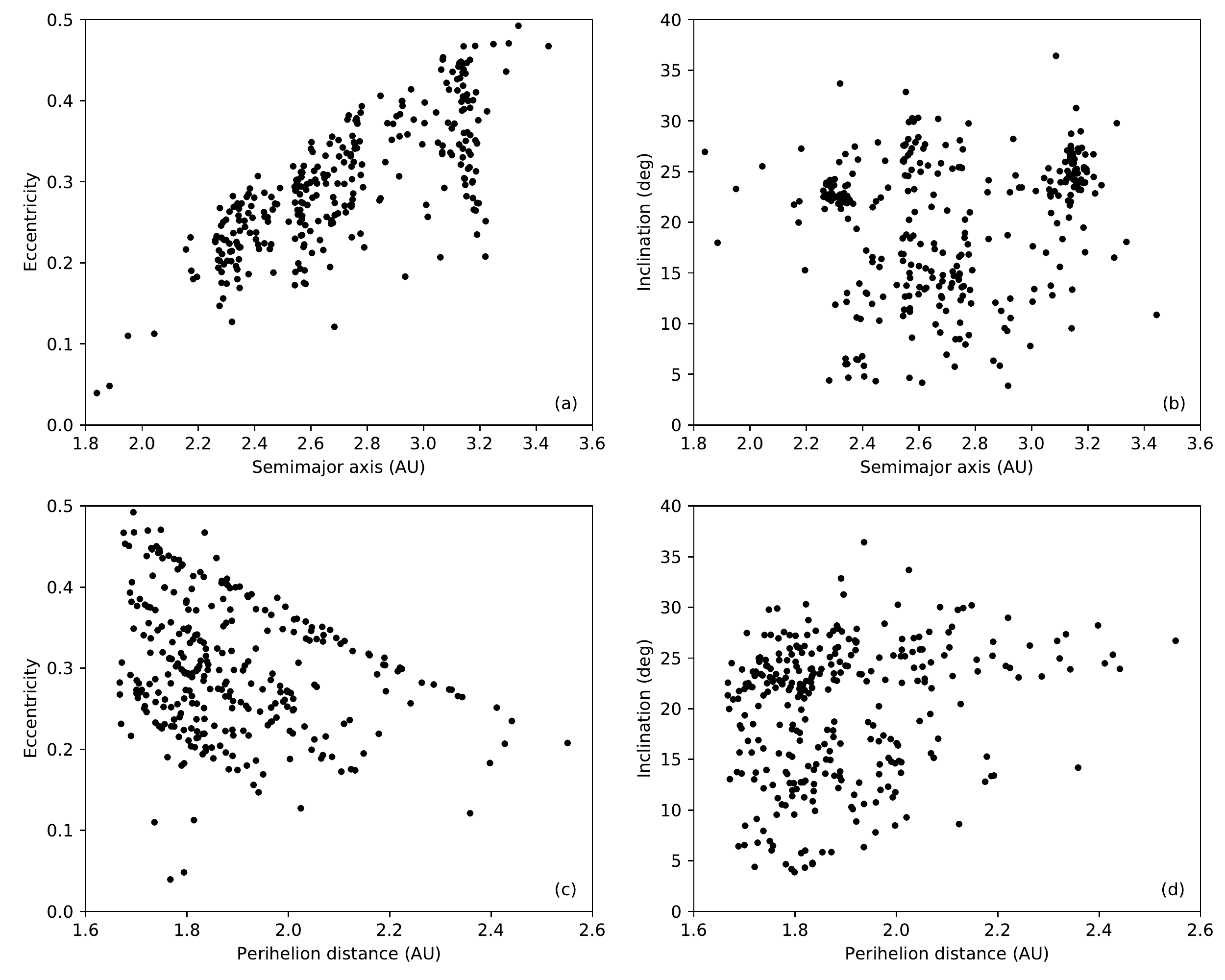}
\protect\caption{Comparison of the semimajor axes, perihelion
  distances, eccentricities, and inclinations of the Main Belt
  asteroids discovered by NEOWISE during the Reactivation survey.
  While the majority of discovered objects have perihelia below 2 AU,
  where they will be warmer and thus easier to detect in the infrared,
  a subset have larger perihelia but also larger orbital
  inclinations.}
\label{fig.disc_aqei}
\end{center}
\end{figure}

\begin{figure}[ht]
\begin{center}
\includegraphics[scale=0.6]{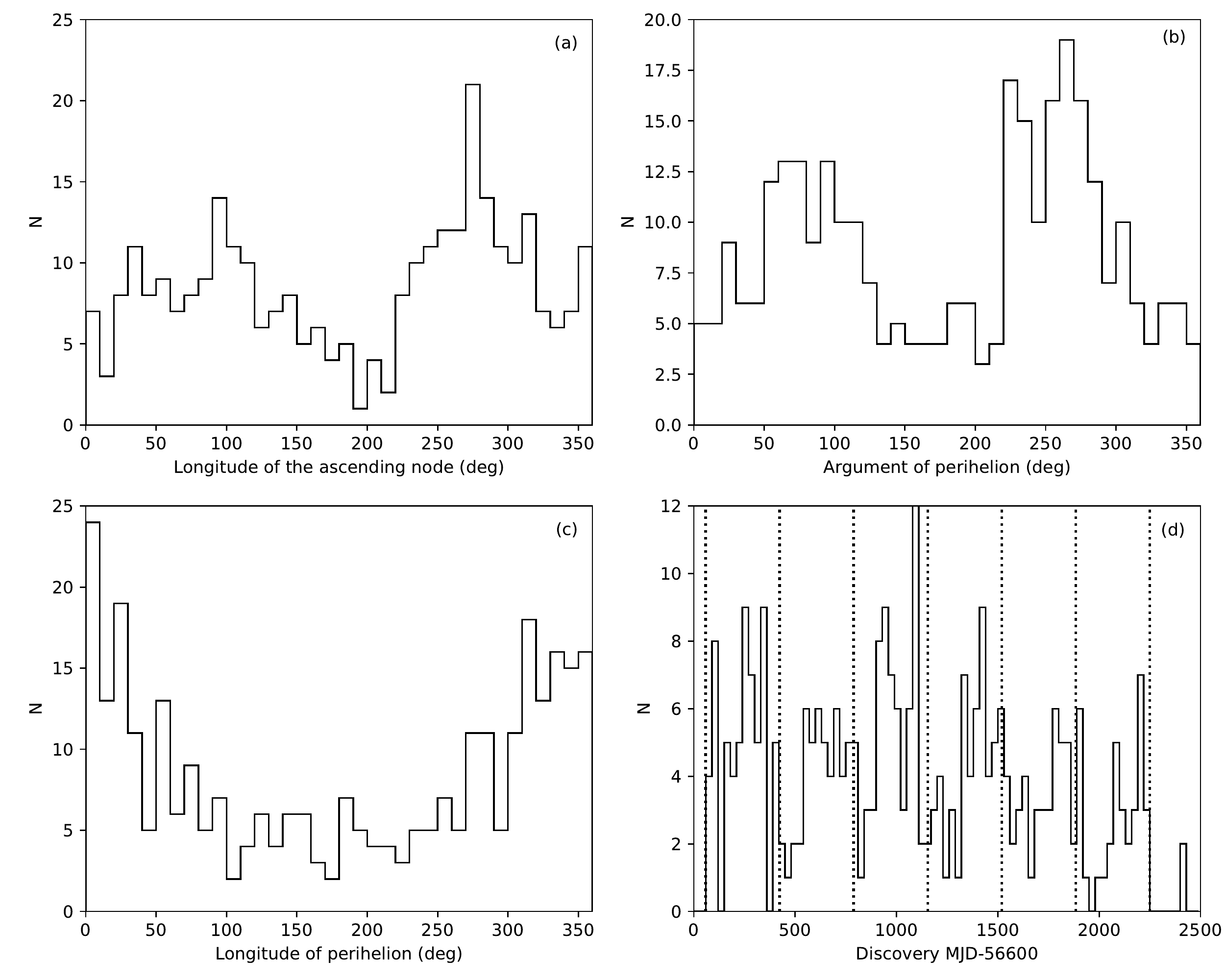}
\protect\caption{Histograms showing the distribution of angular
  orbital parameters and discovery dates for all Main Belt asteroids
  discovered by NEOWISE during the reactivation survey.  While the
  distributions of the longitudes of the ascending node and longitudes
  of perihelion follow the behavior of Main Belt asteroids in general,
  the argument of perihelion distribution shows significant
  differences. Additionally, the date of discovery is skewed toward
  the summer/fall of each year (each Jan 1 is shown as the
  vertical dotted line).  }
\label{fig.disc_hist}
\end{center}
\end{figure}

\begin{figure}[ht]
\begin{center}
\includegraphics[scale=0.5]{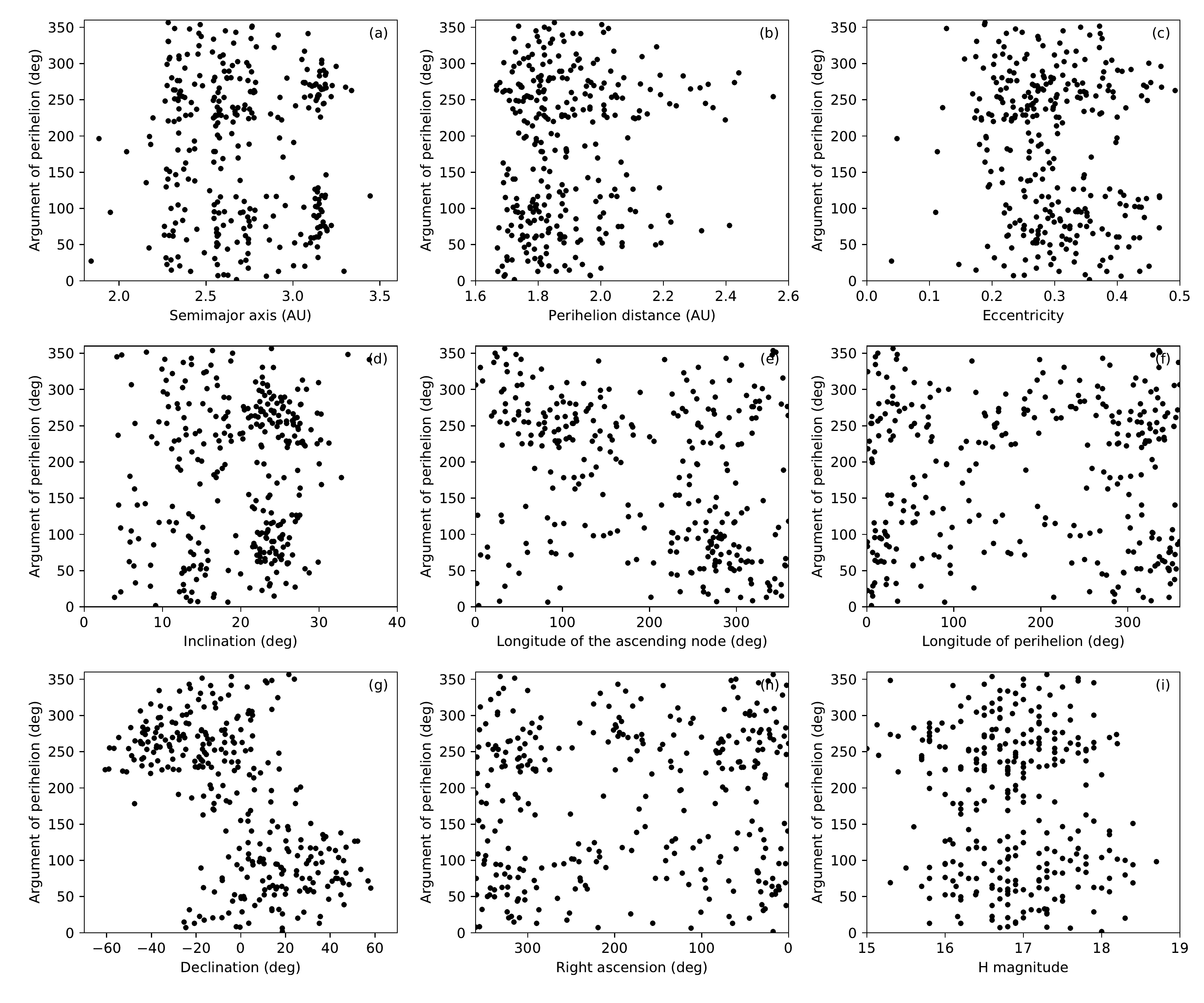}
\protect\caption{Comparison of the argument of perihelion to different
  orbital parameters for all Main Belt asteroids discovered by NEOWISE
  during the Reactivation mission.  Comparisons with perihelion
  distance (b) and declination (g) show the most significant structure.}
\label{fig.disc_om}
\end{center}
\end{figure}

We show in Figures~\ref{fig.disc_aqei}-\ref{fig.disc_om} the
relationship between various orbital parameters for the discovered
MBAs.  Fig~\ref{fig.disc_aqei}(c) shows that the majority of the
discovered objects have perihelion distances ($q$) near the cutoff for
the Mars-crossing population ($q=1.66~$AU). These objects will come
closer to the Sun, have higher surface temperatures and thus be easier
to detect in infrared emission.  Some discoveries, however, have
larger perihelia which makes them less likely to be bright in the
infrared.  These high-q objects tend to have larger inclinations as
shown in Fig~\ref{fig.disc_aqei}(d), and tend to be clumped near the
locations of the Phocaea and Euphrosyne asteroid families
\citep{nesvorny15,masiero15,novakovic17}.

The reason for these correlations begins to be revealed when we look
to the distribution of the angular orbital elements, as shown in
Fig~\ref{fig.disc_hist}. While the structured distribution for the
longitude of the ascending node and the sinusoidal distribution for the
longitude of perihelion match those seen for the broader Main Belt
population\footnote{\it
https://minorplanetcenter.net/iau/lists/MPDistribution.html}, the
argument of perihelion ($\omega$) of the NEOWISE discoveries shows a
strong bimodal distribution that is not seen for the overall
population, which is flat.  The NEOWISE discoveries tend to cluster
around $\omega=90^\circ$ and $\omega=270^\circ$.  Objects with these
orbital parameters have their perihelia at their maximum distance
above/below the ecliptic.  When viewed from Earth, these objects will
be at their closest to the Sun when they are far from the celestial
equator, where ground-based surveys tend to concentrate.

This is further shown in Fig~\ref{fig.disc_om}(g) with objects having
$\omega=90^\circ$ and $\omega=270^\circ$ being discovered at
declinations up to $\pm60^\circ$.  Fig~\ref{fig.disc_om}(b) shows that
the clustering in argument of perihelion for the NEOWISE discoveries
becomes more pronounced at larger perihelion distances.  We interpret
this as indicating that objects missed by ground-based surveys tend to
be those that reach their perihelion far off the ecliptic plane.

Another interesting effect is seen in Fig~\ref{fig.disc_hist}(d),
which shows that NEOWISE Main Belt discoveries tend to occur in the
later half of the calendar year.  We interpret this to as resulting
from a combination of the ecliptic plane crossing the galactic center
near opposition in the summer, the summer monsoon shutdowns
experienced by surveys in Arizona, and the shorter nights for
ground-based telescopes in the summer.  This will cause more objects
to be missed at opposition when they would be most visible to
ground-based telescopes, allowing them to be picked up a few months
later when they enter the NEOWISE scan circle.  The overall decrease
in discoveries with time is likely a combination of increasing
completeness and more recent objects not having had sufficient time to
receive incidental followup, and thus not appearing in the MPC orbital
catalog.  This latter cause is particularly apparent in the low number
of discoveries from 2020.

\clearpage

\section{Conclusions}

We present thermal model fits for near-Earth objects and Main Belt
asteroids observed by the NEOWISE mission during the sixth and seventh
years of its Reactivated mission (13 Dec 2018 to 12 Dec 2020).  These
include diameter and albedo constraints for $227$ epochs of $199$ NEOs
and $6649$ epochs of $5851$ MBAs from Year 6 and $197$ epochs of $175$
NEOs and $6656$ epochs of $5861$ MBAs from Year 7.  When combined with
previously published analyses, during its Reactivation mission NEOWISE
has provided diameter constraints for $1415$ unique NEOs.  Including
fits from the original cryogenic and post-cryogenic mission data
increases this total to $1845$ unique NEOs that have been
characterized.  The NEOs discovered by NEOWISE tend to be low albedo
and a few hundred meters in size, a region of phase space that is not
well-probed by other NEO surveys.

With the larger data set now available, we have compared the NEO
diameter fits based on the W1 and W2 bandpasses to those from the
cryogenic mission phase that allowed for fitting of the beaming
parameter, as well as to independent diameter determinations from
radar observations.  We find that needing to assume a beaming
parameter when only using data significantly shortward of the thermal
peak results in a diameter uncertainty of $\sim30\%$ when compared to
fits using the cryogenic WISE data that span the thermal peak.
Comparisons to independent diameter constraints from radar show a
larger diameter uncertainty that is consistent with a combination of
the need to use a fixed beaming and the deviations of the NEATM model
from actual asteroid thermal behavior, however this is based on the
small sample of overlapping objects.

We also investigated the small number of asteroids in the Main Belt
still being discovered by NEOWISE.  We show that these objects tend to
reside on orbits where their perihelia are significantly out of the
ecliptic plane, resulting in detection at high ecliptic latitudes that
other surveys do not cover as deeply. There also is a correlation in
these discoveries with time of year, which we attribute to a
combination of the effects of the ecliptic plane crossing the galaxy
in the ground-based surveys' field of regard as well as monsoon
seasonal effects for surveys in the southwest US.  For future surveys
that probe fainter magnitudes, a highly complete catalog of MBAs will
be critical to maximizing the capability to discover new NEOs.

The NEOWISE survey is currently on-going, even as the spacecraft's
orbit continues to precess away from its original terminator-following
path. Low levels of Solar activity during the last Solar minimum
resulted in significantly slower orbital evolution, and thus smaller
increases in the focal plane temperature than originally anticipated.
As the Sun leaves Solar minimum and becomes more active orbital
precession is expected to increase, but predictions indicate that the
increase in heating will not have an impact on data quality through at
least mid-2022. Until such time as the orbital evolution results in
significant data degradation, or eventual spacecraft reentry, NEOWISE
provides a unique set of measurements of asteroids and comets passing
close to Earth and is an important component of Earth's planetary
defense ecosystem.

\section*{Acknowledgments}

The authors thank the two anonymous referees for their comments, which
improved the manuscript.  This publication makes use of data products
from the Wide-field Infrared Survey Explorer, which is a joint project
of the University of California, Los Angeles, and the Jet Propulsion
Laboratory/California Institute of Technology, funded by the National
Aeronautics and Space Administration.  This publication also makes use
of data products from NEOWISE, which is a joint project of the
University of Arizona and the Jet Propulsion Laboratory/California
Institute of Technology, funded by the Planetary Science Division of
the National Aeronautics and Space Administration.  This research has
made use of data and services provided by the International
Astronomical Union's Minor Planet Center.  This research has made use
of the NASA/IPAC Infrared Science Archive, which is funded by the
National Aeronautics and Space Administration and operated by the
California Institute of Technology.  This research has made use of
NASA’s Astrophysics Data System. This research has made extensive use
of the {\it numpy}, {\it scipy}, and {\it matplotlib} Python
packages. The authors also acknowledge the efforts of worldwide NEO
followup observers, both professional and non-professional, who
provide time-critical astrometric measurements of newly discovered
NEOs enabling object recovery and computation of orbital elements.
Many of these efforts would not be possible without the financial
support of the NASA Planetary Defense Coordination Office, for which
we are grateful.

\end{document}